# The Circular Atwood Machine

Martín Monteiro[a], Cecilia Stari[b], Arturo C. Marti[c],
[a] Universidad ORT Uruguay; monteiro@ort.edu.uy
[b] Universidad de la República, Uruguay, cstari@fing.edu.uy
[c] Universidad de la República, Uruguay, marti@fisica.edu.uy

**Abstract**: The Atwood Machine, a classic apparatus in physics education, has historically been pivotal in demonstrating Newtonian mechanics, specifically Newton's Second Law. This study introduces an innovative adaptation, the circular Atwood machine, aimed at exploring circular motion and angular dynamics and integrating emerging technologies, specifically smartphone sensors. Through a rotating disc subjected to controlled external torque, the experiment delves into the relationship between the torque applied and the resulting angular momentum. The study not only presents the theoretical framework but also outlines a practical setup using readily available materials, emphasizing the potential for contemporary technology to enhance the comprehension and teaching of fundamental physical concepts.

## 1. Introduction

The Atwood Machine, a traditional device in physics, has been an invaluable tool for demonstrating and understanding the fundamental principles of Newtonian mechanics, especially Newton Second Law[1]. Its application in education has been extensively documented over the years (see for example[2]). In relation to the device itself, it is possible to find Atwood machines ranging from very simple ones to high quality handcrafted devices worthy of being exhibited in museums. Some time ago, we have published in this column a proposal which makes use of the smartphones and their built-in sensors to propose an updated version of the experiment but keeping its essence unchanged[3]. Now, we propose a *circular* Atwood machine designed to study the relationship between the external torque on a rigid and its angular acceleration mediated through the inertia moment. In our proposal, quantitative measurements are achieved by means of a smartphone's angular velocity sensor. The inertia moment measured was successfully contrasted with the value obtained from the period of oscillations of the device hanging from a point at its periphery.

## 2. The circular Atwood machines

The traditional Atwood machine allows to analyze linear forces and acceleration. However, by adapting it to investigate circular motion, we open up new opportunities for exploring more delicate concepts in physics, such as circular kinematics and angular dynamics. The essence of the circular Atwood's machine is a rotating disc subjected to an external torque in which the inertia moment and the torque can be controlled by the experimenter. The device also allows us to easily measure the angular acceleration, which is used to relate the external torque to the angular momentum of the disk. Figure 1 (left panel) shows the top and lateral view of the circular Atwood's machine and a photograph (right panel).

Let us analyze the dynamics of the circular Atwood machine. The device is composed of a disc that can rotate around the vertical axis subject to an external torque given by a thread with tension $T$ by means of a pulley of radius $R$. In general, we must take the rolling resistance in the bearings $\tau_{roz}$ into account. We denote $I$ the device's moment of inertia about a vertical axis through the center of mass, and $\alpha$ the angular acceleration. We recall that, as a consequence of Newton's laws applied to a rigid body, the total external torque is equal to the moment of inertia times the angular acceleration. Given that we are interested in these magnitudes projected on the vertical axis. Considering the vertical axis and the total external torque is the sum of the torque exerted by the pulley minus the rolling friction, we can write

$$TR - \tau_{roz} = I\alpha \qquad (1)$$

On the other hand, the tension $T$ can be determined by applying Newton second law to the mass hanging from the device along the same axis

$$m_0 g - T = m_0 a \qquad (2)$$

where $a$ is the vertical acceleration of the mass and, since the string does not slip with respect to the pulley can be related to the angular acceleration of the disk as

$$a = \alpha R \qquad (3)$$

The moment of inertia of the device can be varied by symmetrically placing the two masses shown in Fig. 1 at different distances $r$ from the center of rotation. For simplicity, we write the moment of inertia as the sum of the device's moment of inertial $I_D$, which remains constant, and the (variable) contributions of the masses

$$I = I_D + 2mr^2 \qquad (4)$$

From the previous equations we can eliminate $T$ and $a$ to obtain

$$\frac{1}{\alpha} = \frac{I_D + 2mr^2 + m_0 R^2}{m_0 g R - \tau_{roz}} \qquad (5)$$

expressing a relationship between the inverse of the angular acceleration and the distance $r$. In principle, it is possible to design different experiments with the circular Atwood machine. In this opportunity we record the angular acceleration values as a function of $r$; i.e. placing the masses in the different slots of the support.

3. The experiment

The experimental device, shown in Fig. 1 (right panel), was built from an old record player and includes a 3D printed plastic part, masses and spring. The plastic piece is a support where the smartphone can be hold, and has slots to place the masses, $m=0.0521(5)$ kg each one, and is attached to the center of the tray. The design was made in Tinkercad and is available for anyone interested to copy and modify to suit their own experiment[4]. The slots are equispaced and located on a diameter so as to vary the distance to the center of rotation, $r$, in a regular way. The disc is connected to a hanging mass, $m_0=0.1265(5)$ kg, by means of pulley with radius $R=0.0259(1)$m. . The weights of the strings and the pulley can be neglected.

A smartphone Nexus 5 is placed on the holder to register the angular velocity by means of the gyroscope (rotation sensor) included in the built-in MPU6515 MEMS chip. Although many experiments using smartphone sensors have been proposed in recent years[5], the rotation sensor has been less used (see for example[6-9]). In this experiment we use the *phyphox* app[10] to record the angular velocity. For each location of the masses we record the angular velocity and, as shown in the screenshot of Fig. 2, we identify the falling interval of the hanging mass and calculate the slope of the graph corresponding to the angular acceleration. For each value of the distance *r* we measure the angular acceleration. The measurements are shown in table I.

| r (m) | α (rad/s²) |
|---|---|
| δr=0.0001 m | δα=0.01 rad/s² |
| 0.0900 | 12.29 |
| 0.0800 | 13.20 |
| 0.0700 | 14.16 |
| 0.0600 | 14.78 |
| 0.0500 | 16.06 |
| 0.0400 | 16.87 |
| 0.0300 | 17.77 |
| 0.0200 | 18.33 |

Table I. Measurements of distance of the masses and angular acceleration.

Figure 3 summarizes the experimental results. Taking into consideration Eq. (5) we choose to plot the inverse of the angular acceleration as a function of the distance r squared. We performed a linear regression of the form $\frac{1}{\alpha}=ar^2+b$ where the slope, *a*, and the ordinate at the origin, *b*, are easily related to the parameters of Eq. (5)

$$a=\frac{2m}{m_0 gR - \tau_{roz}} \quad \text{and} \quad b=\frac{I_D + m_0 R^2}{m_0 gR - \tau_{roz}} \tag{6}$$

obtaining a linear correlation coefficient of r = 0.9974953. From these equations we obtained the values of the friction torque and the inertia moment of the device as $\tau_{roz}=0,00232(2)$ N.m and $I_D=1.48(2)\times 10^{-3}$ kg.m².

## 4. A direct measure of the inertia moment

To obtain the inertia moment the device was attached to a point on the periphery and allowed to oscillate in a vertical plane as indicated in Fig. 4. To measure the period we again used the rotation sensor with the help of the *phyphox* app, in this case the period obtained by averaging over 10 oscillations was T=0.735(2). The total mass of the device was $M_D$=0.3770(5) kg (including disc, pulley, support and smartphone) and the distance from the point of oscillation to the center of rotation was d=0.0910(2) m. Applying Steiner parallel axis theorem to the period of oscillation of the system,

$$T=2\pi\sqrt{\frac{I_D + M_D d^2}{M_D g d}} \tag{7}$$

we find $I_D=1.48(6)\times 10^{-3}$ kg.m², in great concordance with the value obtained from the angular acceleration fit.

## 5. Discussion and conclusion

We presented an innovative approach that leverages these emerging technologies to transform the Atwood Machine into a tool for studying circular dynamics. This experiment can be set up using a commercial device or built using a tray from an old record player plus a holder (which can be 3D printed[4]), some bearings and other readily available materials. Using the angular velocity sensor of a smartphone it is possible to measure the angular acceleration in a simple way and link it to the angular momentum of the device. The relationship can be linearized and the adjusted parameters can be linked to the friction at the bearing and the inertial moment of the device which can be independently measured. This experiment represents an exciting step forward in the application of contemporary technologies to comprehend and teach fundamental physical concepts.

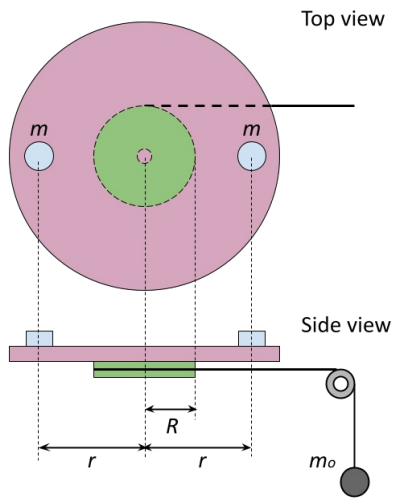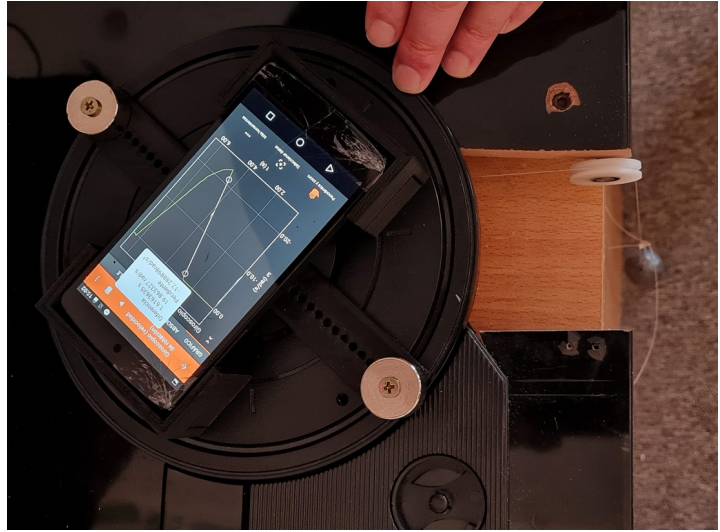

**Figure 1.** Experimental setup: schematic representation (left panel) and photograph (right panel).

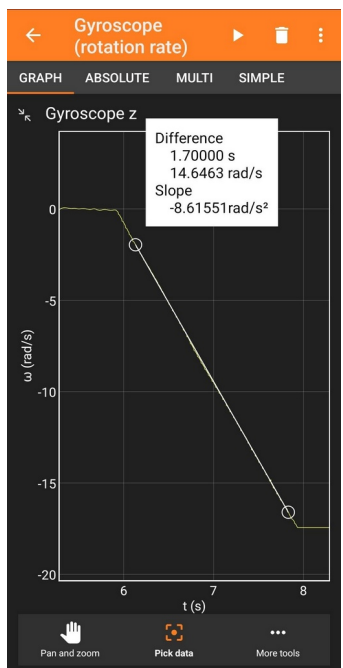

**Figure 2.** Screenshot of the phyphox app used to register angular velocity along the vertical axis. The angular acceleration is obtained by means of the slope of the graph as indicated.

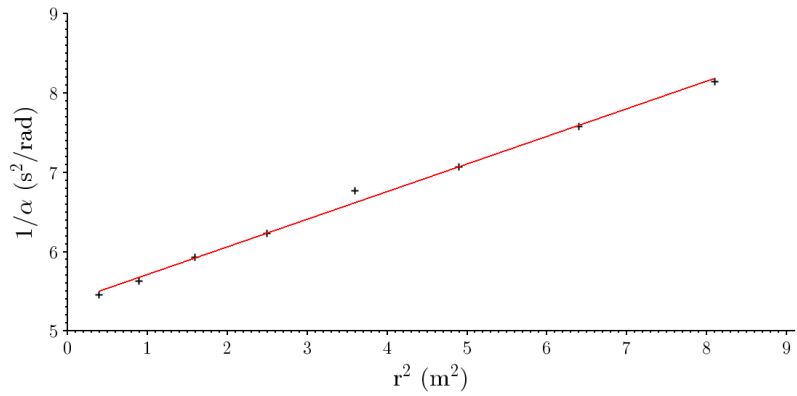

**Figure 3.** Linearized relationship between the angular velocity and the location of the masses. Experimental results (crosses) and linear fit (red line).

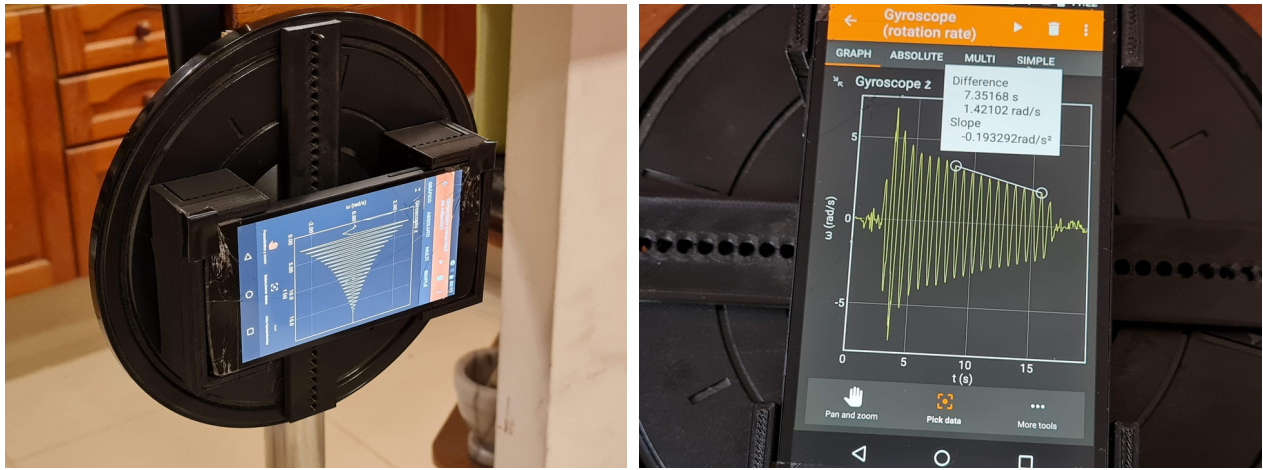

**Figure 4.** To directly measure the inertia moment the device was hung from a point at its periphery, let it oscillate, and the period of oscillation was registered using the *phyphox* app (left picture). The period was obtained by the time difference tool over 10 oscillations (right picture).